\documentclass[12pt]{iopart} 
\usepackage{graphicx} 
\usepackage{amssymb}
\begin{document}

\title{  Anisotropic susceptibility of the geometrically frustrated spin-chain
compound Ca$_3$Co$_2$O$_6$ }

\author{ Vincent Hardy, Delphine Flahaut, Raymond Fr\'{e}sard and Antoine
  Maignan}

\address{ Laboratoire CRISMAT, UMR CNRS--ENSICAEN(ISMRA) 6508, \\  
             6 Bld. du Mar\'echal Juin, F-14050 Caen, France}  

\ead{Vincent.Hardy@ensicaen.fr}

\begin{abstract}
Ca$_3$Co$_2$O$_6$ is a system exhibiting a series of fascinating properties,
including magnetization plateaus and remarkably slow dynamics at low-$T$.
These properties are intimately related to the geometrical frustration,
which results from a particular combination of features: (i) the chains are
arranged on a triangular lattice; (ii) there is a large uniaxial anisotropy;
(iii) the intrachain and interchain couplings are ferromagnetic and
antiferromagnetic, respectively.

The uniaxial anisotropy is thus an issue of crucial importance for the
analysis of the physical properties of Ca$_3$Co$_2$O$_6$. However, it turns
out that no precise investigation of this magnetic anisotropy has been
performed so far. On the basis of susceptibility data directly recorded on
single crystals, the present study reports on quantitative information about
the anisotropy of Ca$_3$Co$_2$O$_6$.
\end{abstract}


\submitto{\JPCM}  

\maketitle

\section{Introduction}

\begin{figure}[b!]
\begin{center}
\DeclareGraphicsExtensions{.eps,.eps.gz,{}}
\includegraphics*[width=0.5\textwidth ]{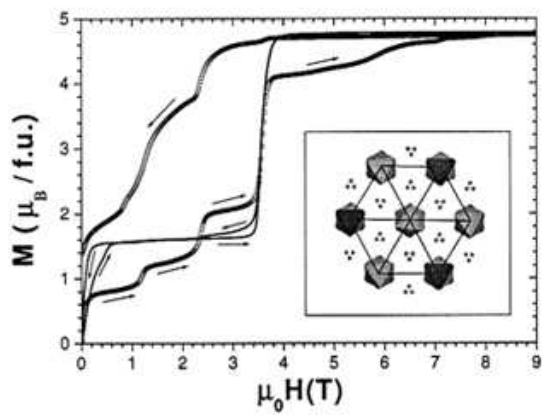}
\end{center}
\caption{ 
Hysteresis loops ($H\parallel c$) recorded with a sweep rate of 0.1
T / min. at 2 K (circles) and 10 K (solid line). The arrows indicate the
direction of the field variation. The inset shows a projection of the
structure along\ the hexagonal $c$-axis (the dark and light polyhedra
represent CoO$_6$ trigonal prisms and CoO$_6$ octahedra, respectively; the
grey circles represent the Ca ions; the solid lines emphasize the triangular
arrangement of the chains in the $ab$ plane).  }
\label{fig:1}
\end{figure}

\noindent Ca$_3$Co$_2$O$_6$ is a geometrically frustrated spin-chain
compound which has attracted considerable attention in recent years, owing
to a series of puzzling properties \cite
{FJ96,AA97,KA97a,KA97b,MA00,CpSet4,CpCeram,RAY03b,FRE04,HAR04a,FLA04,HAR04b}.
The structure of this compound consists of chains made up of CoO$_6$
trigonal prisms alternating with CoO$_6$ octahedra, which run along the $c$%
-axis of the hexagonal cell \cite{FJ96}. These chains are separated by the Ca
ions and they form an hexagonal lattice on the $ab$ plane (see inset
of Fig.~\ref{fig:1}). The issue of the valence and spin states of the
Co ions on these two 
sites has been the subject of intense controversies. Owing to a series of
recent results \cite{WHO03,SAM04,EY04,TA05,WU05}, it is now widely accepted
that (i) the Co ions are trivalent for both sites; (ii) owing to the
difference in the crystalline electric field (CEF), the Co$^{3+}$($3d^6$)
ions are in the high-spin state for the prismatic sites ($S=2$) whereas they
are in the low-spin state ($S=0$) for the octahedral sites. In other
respects, it was early found that the intrachain coupling $J$ is
ferromagnetic while the interchain coupling $J^{\prime }$ is
antiferromagnetic \cite{AA97}. Furthermore, Ca$_3$Co$_2$O$_6$ was found to
exhibit strongly anisotropic magnetic properties, the spins having a
preferential orientation along the $c$ axis \cite{KA97b,MA00}. Most probably,
this last feature can be ascribed to the single-ion (uniaxial) anisotropy of
the $S=2$ spins at the prismatic sites.

Ca$_3$Co$_2$O$_6$ possesses a combination of features (i.e. triangular
lattice with spins oriented along the chain direction and with an
antiferromagnetic interchain coupling) which yields a situation of
geometrical frustration \cite{CO97}. It is clear that this frustration plays
a great role in the peculiar magnetic properties of Ca$_3$Co$_2$O$_6$. For
instance, the long-range ordering which takes place below $T_N\simeq 26$ K
\cite{AA97} was found to be closely related to the ``Partially Disordered
Antiferromagnetic'' state (PDA) \cite{KA97a} proposed by Mekata \cite{PDA}
about the ABX$_3$ compounds, another family of geometrically frustrated
spin-chains \cite{CO97}. In the PDA\ state, two over three chains are
antiferromagnetically coupled (with long-range order along each of them),
while the third one is left incoherent (i.e. without long-range order along
its direction and no correlation with its neighbors). Among the peculiar
magnetic properties of Ca$_3$Co$_2$O$_6$, one can mention (i) the appearance
of a spin freezing at $T_{SF}\,(=T_{f=0})\sim 8$ K \cite
{KA97a,CpSet4,CpCeram}, leading to a pronounced frequency dependence of the
susceptibility ($\Delta T_f/T_f)/\Delta \log f\simeq 0.17$) \cite
{MA00,HAR04b}; (ii) for $T<T_{SF}$, the existence of magnetization steps with
a roughly constant field spacing \cite{KA97b,MA00,QTMnous} and the
appearance of a saturation in the spin-relaxation time, these features
having led us to suggest the possibility of a phenomenon of Quantum
Tunneling of the Magnetization (QTM) \cite{QTMnous,HAR04b}.

It can be emphasized that the geometrical frustration in Ca$_3$Co$_2$O$_6$
is strongly related to the existence of a pronounced uniaxial anisotropy. In
spite of this, we observe that no precise quantitative analysis of this
magnetic anisotropy has been performed so far. It is the goal of the present
study to extract quantitative information about the anisotropy of Ca$_3$Co$%
_2 $O$_6$.

\section{Experimental details}

Single crystals of Ca$_3$Co$_2$O$_6$ were grown according to the following
method: a mixture of Ca$_3$Co$_4$O$_9$ and K$_2$CO$_3$, in a weight ratio of
1/7, was heated at 950 ${{}^{\circ }}$C for 50 hours in an alumina crucible
in air; then, the cooling was performed in two steps, first down to 930 ${%
{}^{\circ }}$C at 10 ${{}^{\circ }}$C/h and then down to room temperature at
100 ${{}^{\circ }}$C/h. This procedure leads to crystals having a
needle-like shape (with the $c$ axis along the longest dimension) which is
convenient for orientation purposes. These crystals can be quite long
(typically 4 mm) but they are thin (less than 0.5 mm). Accordingly, we
performed the measurements on an assembly of crystals. Owing to their shape,
we emphasize that the alignment between them, as well as with respect to the
field direction (i.e. $c$ axis either parallel or perpendicular to the field
direction) can be precisely achieved. Moreover, the measurements for both
orientations were carried out with the same mounting system ---just by
rotating the set of crystals inside the sample rod--- in order to keep
constant the residual background signal.

\begin{figure}[b!]
\begin{center}
\includegraphics*[width=0.7\textwidth ]{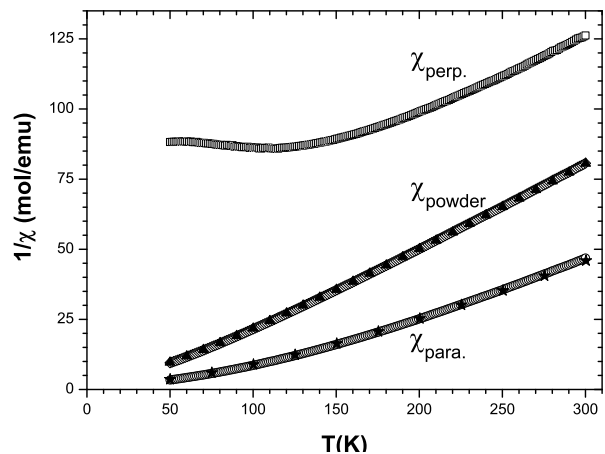}
\end{center}
\caption{ 
Susceptibility data recorded on single crystals of Ca$_3$Co$_2$O$_6$%
, with the magnetic field applied either along the $c$-axis ($\chi _{para}$
referred to as $\chi _{\parallel }$ in the text) or perpendicular to the
c-axis ($\chi _{perp}$ referred to as $\chi _{\perp }$ in the text)$.$ Also
shown is the corresponding ``powder'' data derived from the general
expression $\chi _p=$($\chi _{\parallel }+$ $2\chi _{\perp })/3$. The small
filled triangles represent the data directly recorded on a ceramic sample.
The small filled stars represent the data previously recorded on another set
of crystals (different morphology with a larger size) with $H\parallel c$. }
\label{fig:2}
\end{figure}

Curves of magnetization as a function of $T$ were recorded in 0.1 T, for
both orientations, by using a Superconducting Quantum Interference Device
magnetometer (MPMS, Quantum Design). We checked that, with 0.1 T,\ we are in
the linear regime of the $M(H)$ curves for both orientations and over the
whole investigated temperature range, $50<T<300$ K (i.e. $T>>T_N$). In what
follows, one considers the susceptibility curves obtained by dividing the
magnetization by the measuring field, leading to $\chi _{\parallel }(T)$
(i.e. with $H\parallel c$) and $\chi _{\perp }(T)$ (i.e. with $H\perp c$).
Two observations support the reliability of these data: first, we found that
the $\chi _{\parallel }(T)$ is in perfect agreement with that previously
obtained using bigger crystals (of different shape); second, the powder-like
data obtained by calculating $\chi _p=(\chi _{\parallel }+$ $2\chi _{\perp
})/3$ is found to be superimposed on the susceptibility curve registered on
a ceramic (see Fig.~\ref{fig:2}).

\section{Results and Discussions}

Figure~\ref{fig:2} shows the $\chi _{\parallel }(T)$\ and $\chi
_{\perp }(T)$ data 
---as well as the derived $\chi _p(T)$ curve--- along with data recorded on
a ceramic sample and on bigger crystals with $H\parallel c$. These data well
exhibit the considerable magnetic anisotropy of this compound. To be more
quantitative, however, one needs a proper modelization of the
susceptibilities. It turns out that it is far from being obvious in the case
of Ca$_3$Co$_2$O$_6$ which gathers several particularities.

As previously discussed, we can assume that the anisotropy of Ca$_3$Co$_2$O$%
_6$ is basically a single-ion feature, originating from a combination of the
CEF and spin-orbit effects. We can also simply consider a Heisenberg form
for the intrachain coupling, while the interchain coupling (much weaker) can
be neglected in a first approximation. In other respects, we note that a
spin chain compound would require in principle a specific treatment of the
magnetic coupling in order to account for the enhanced fluctuation effects
in 1D systems.

To the best of our knowledge, there is no model in the literature which
takes into account all the features of Ca$_3$Co$_2$O$_6$. In what follows,
we will consider two modelizations which can be applicable ---to some
extent--- to the case of Ca$_3$Co$_2$O$_6$. We emphasize that each of these
models only accounts for a part of the features of Ca$_3$Co$_2$O$_6$. It
also deserves to be noted that they correspond to different approaches that
can be complementary in the analysis of magnetism.

\subsection{First model}

We found in the literature a model which made a precise treatment of the
on-site magnetism in a situation very close to ours. This work carried out
by Parkin and Friend \cite{PA80} dealt with iron intercalates of
dichalcogenides, with Fe$^{2+}$ (i.e., 3d$^6$ like Co$^{3+}$) in a
trigonally distorted anionic environment. It turns out that the cubic part
of the CEF splits the $^5D$\ ground state, and generates a $\Gamma _5$
triplet at lower energy, which can be represented by a fictitious orbital
angular momentum $\widetilde{{\bf L}}$. On this basis, Parkin and
Friend \cite{PA80} performed a rigorous analysis by considering the following
single-site Hamiltonian:

\begin{equation}
H_i=-k_B\delta \widetilde{L}_{z,i}^2-k_B\lambda \,\widetilde{{\bf L}}_i\cdot 
{\bf S}_i+\mu _B(-\widetilde{{\bf L}}_i+2{\bf S}_i)\,{\bf B\;.}
\end{equation}

\smallskip The first term corresponds to the effect of the non-cubic part of
the CEF, the second term is the spin-orbit (SO) coupling, and the last term
is the basic form of the Zeeman energy. For our compound with such an
Hamiltonian, $\delta $ is positive while $\lambda $ is negative. There is no
magnetic coupling in this model but we note that a spin-spin interaction can
be added ---if necessary--- by using a mean-field (MF) approximation (even
though we know such a MF treatment is poorly suited to 1D\ systems).

Using the eigenfunctions and associated energies, Parkin and Friend
calculated the susceptibilities using the Van Vleck formula. Extending their
approach, we took into account all the levels of the lower triplet (whereas
only the two lowest are considered in Ref.~\cite{PA80}), but it actually
makes almost no difference. The expressions found for the susceptibilities
can be written as:

\begin{equation}
\chi _{\parallel }=\frac{N\,\mu _B^2\,S(S+1)}{3k_B}\;\frac 1T\;\left[ \frac{%
f_{\parallel }}Z\right]
\end{equation}

\begin{equation}
\chi _{\perp }=\frac{N\,\mu _B^2\,S(S+1)}{3k_B}\;\frac 1T\;\left[ \frac{%
f_{\perp }}Z\right]
\end{equation}

where

\begin{eqnarray}
f_{\parallel } &=&25\exp (-2s)+9\exp (-s)+1+\exp (s)+ \\
&&9\exp (2s)+16\exp (\frac{2s^2}p+p)+4\exp (\frac{5s^2}p+p)\;,  \nonumber
\end{eqnarray}

\begin{eqnarray}
f_{\perp } &=&(-8/s)\exp (-2s)-(4/s)\exp (-s)+(4/s)\exp (s)+ \\
&&(8/s)\exp (2s)+16\exp (\frac{2s^2}p+p)+4\exp (\frac{5s^2}p+p)\;,  \nonumber
\end{eqnarray}

\begin{eqnarray}
Z &=&2[\exp (-2s)+\exp (-s)+1+\exp (s)+\exp (2s)+ \\
&&\exp (\frac{2s^2}p+p)+\exp (\frac{5s^2}p+p)+1/2\exp (\frac{6s^2}p+p)]\;, 
\nonumber
\end{eqnarray}

with $s=\lambda /T$ and $p=-\delta /T$. All others parameters have their
usual meaning.

It can be emphasized that the exact value of $\delta $ has only a tiny
influence on the final results (we used $\delta =-100$ K hereafter). Roughly
speaking, one can consider that the main role of the non cubic part of the
CEF is to split the $\Gamma _5$ triplet, which allows the SO to generate
anisotropy. In other respects, it must be realized that $\lambda $ cannot
really be regarded as a free parameter in these expressions, since the spin
orbit parameter is supposed to be an intrinsic characteristic of each ion,
not expected to vary significantly with the environment. In the case of Co$%
^{3+}$, $\lambda $ must be close to -145 cm$^{-1}$ \cite{AB70}. Accordingly,
it deserves to be noted that there is almost no adjustable parameter in this
model, apart from a possible correction related to $J$.

\begin{figure}[h!]
\begin{center}
\includegraphics*[width=0.7\textwidth ]{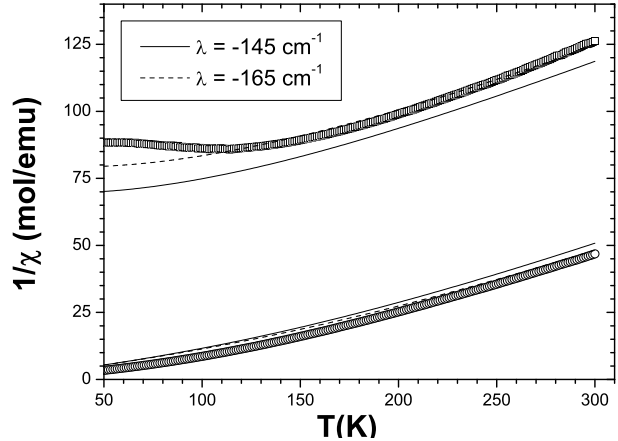}
\end{center}
\caption{ Experimental $\chi _{\parallel }(T)$\ and $\chi _{\perp }(T)$
curves, together with the calculations using the model of Parkin and Friend
with $\lambda =-145$ cm$^{-1}$ (solid lines) and $\lambda =-165$ cm$^{-1}$
(dashed lines)$.$ }
\label{fig:3}
\end{figure}

Figure~\ref{fig:3} shows the raw data, along with the curves
calculated for two values 
of $\lambda $. With the expected value $\lambda =-145$ cm$^{-1}$, the
calculated curves do not lie far from the experimental ones, they have the
right shape in the high-$T$ range, but there is a sizeable shift from the
data. Moreover, the model cannot account for the existence of a minimum in $%
1/\chi _{\perp }(T)$, as it is visible in the data around 120 K. We also
observed that adding a correction related to the intrachain coupling can
yield a very nice fitting for $\chi _{\parallel }(T)$\ but not for $\chi
_{\perp }(T)$. Strikingly, it can be noticed that remarkable fittings of
both curves can be obtained with $\lambda =-165$ cm$^{-1}$. As previously
discussed, however, such a significant departure from the free ion value of $%
\lambda $ is suspect. In conclusion, one must state that this model cannot
well describe the magnetic anisotropy of Ca$_3$Co$_2$O$_6$, even though it
well illustrates the leading role of the spin orbit coupling.

\subsection{Second model}

Let us now consider a very different approach, approximating the
susceptibility in the form of series expansion. This general method discards
the specificities of the system under investigation (in contrast to the
previous model), but it was found to be powerful in many cases, even though
it is by definition limited to the high-$T$ part of the data (i.e.
applicable only for $kT$ larger than all the energy terms relevant to the
system). Yosida \cite{YO51} has derived such a series expansion which appears
to be suitable to the case of Ca$_3$Co$_2$O$_6$, in the sense that it
accounts for both a magnetic coupling J and an anisotropy parameter D. The
starting point is an Hamiltonian of the form:

\begin{equation}
H_i=-k_BDS_{z,i}^2-2k_BJ\sum_{j\neq i}{\bf S}_i{\bf S}_j+g\mu _B{\bf S}_i%
{\bf B\;,}
\end{equation}

where the index $j$ refers to the first neighboring spins. In such a
picture, the Zeeman term involves a Land\'{e} factor (which can be an
adjustable parameter), while the uniaxial anisotropy appears as an effective
term applied to the spin. Even though it can carry some ambiguities, this
form of Hamiltonian is the most common to deal with uniaxial anisotropy, and
it is very useful in practice. We underline that the $\delta $ parameter of
the previous model strongly differs from $D$ in the above equation (this
latter parameter incorporating both the CEF\ and SO effects).

The formula derived by Yosida can be written as follows:

\begin{equation}
\chi _{\parallel }=\frac C{T-\theta }[1+2\frac QT]\;,
\end{equation}

\begin{equation}
\chi _{\perp }=\frac C{T-\theta }[1-\frac QT]\;,
\end{equation}

In these expressions,

- $C$\ is the usual Curie constant

\begin{equation}
C=\frac{Ng^2\mu _B^2S(S+1)}{3k_B}
\end{equation}

- $\theta $ is the characteristic temperature associated to $J$ (i.e. the
Curie-Weiss temperature)

\begin{equation}
\theta =\frac{2zJS(S+1)}3
\end{equation}

(here $z$ is the number of nearest neighbors, i.e. 2 in the case of
intrachain coupling)

- $Q$ is a characteristic temperature associated to $D$

\begin{equation}
Q=D\;\left[ \frac{2S(S+1)}{15}-\frac 1{10}\right]
\end{equation}

Note that, in our case, $\theta >0$ and $Q>0$.

\smallskip In Eqs (8-9), they are three free parameters which should be
adjusted to best fit to both the $\chi _{\parallel }(T)$\ and $\chi _{\perp
}(T)$ data. Actually, one can also use combinations of $\chi _{\parallel }(T)
$\ and $\chi _{\perp }(T)$ allowing to isolate the influence of some of
these parameters. For instance, it can be noted that the parameter $D$
disappears in the powder formula, yielding the usual form

\begin{equation}
\chi _p=\frac{\chi _{\parallel }+2\chi _{\perp }}3=\frac C{T-\theta }\;.
\end{equation}

\begin{figure}[t!]
\begin{center}
\includegraphics*[width=0.7\textwidth ]{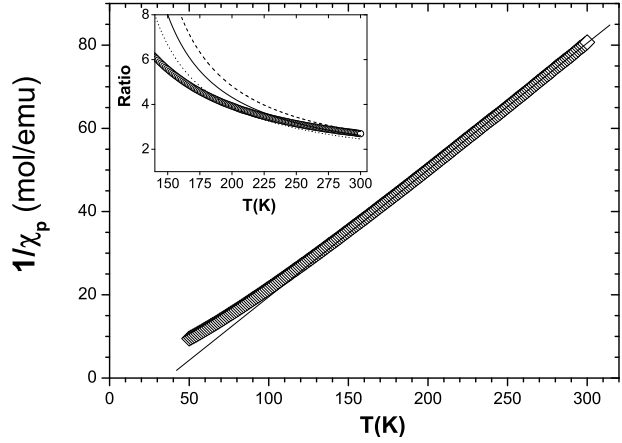}
\end{center}
\caption{ The main panel shows the $1/\chi _p(T)$ curve derived from the data
(see text), which exhibits a linear behavior at high temperature (solid
line). The inset shows the temperature dependence of the susceptibility
ratio ($\chi _{\parallel }/\chi _{\perp })$, together with the expectations
of the Yosida's model for several values of $D$: from bottom to top, $D=140$
K (dotted line), $D=150$ K (solid line) and $D=160$ K (dashed line).}
\label{fig:4}
\end{figure}

As shown in Fig.~\ref{fig:4}, such a behavior is found to be well
obeyed by the data. 
Fitting to the data in the range $200-300$ K leads to $g\simeq 2.08$ and $%
J\simeq 4.5$ K.

Furthermore, the ratio($\chi _{\parallel }/\chi _{\perp })$ is expected to
only depend on $D$

\begin{equation}
\frac{\chi _{\parallel }}{\chi _{\perp }}=\frac{1+2\frac QT}{1-\frac QT}
\end{equation}

Inset of Fig.~\ref{fig:4} shows that there is a restricted range of $D$ values
---around $150$ K--- which leads to a reasonable fitting in the high-$T$
range.

\begin{figure}[t!]
\begin{center}
\includegraphics*[width=0.7\textwidth ]{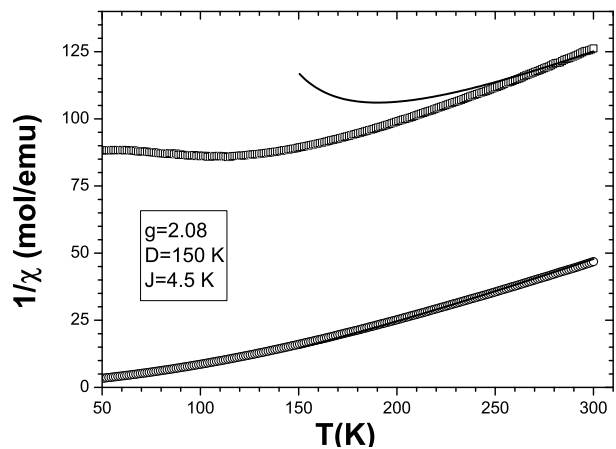}
\end{center}
\caption{ Experimental $\chi _{\parallel }(T)$\ and $\chi _{\perp }(T)$
curves, together with the calculations using the Yosida's model for the
displayed values of parameters (according to the assumptions of this model,
the calculations are limited to the high-$T$ range).}
\label{fig:5}
\end{figure}

The $\chi _{\parallel }(T)$\ and $\chi _{\perp }(T)$ curves, corresponding
to $g=2.08,$ $D=150$ K and $J=4.5$ K, are reported in
Fig.~\ref{fig:5}. The agreement with the data is far from being
perfect, but it is reasonable if one 
considers that the same set of parameters must fit to two curves (not to
mention that perfect fittings are obtained if one authorizes different sets
of parameters for each curve). Probably, a part of the deviation from the
data can be attributed to the fact that the temperatures of the fitting
range are not much larger that $D$. Moreover, this model provides a rough
description of the magnetic coupling since the 1D character is only taken
into account through the value of $z$ (= 2).

\section{Conclusion}

The anisotropic susceptibility of the geometrically frustrated spin chain Ca$%
_3$Co$_2$O$_6$ has been measured on single crystals. Two models of the
literature have been used to analyze these data. None of them was found to
be perfectly satisfying. However, the first one clearly pointed to the role
of the SO in this anisotropy \cite{PA80}, while the second one provided us
with a reliable estimate of the effective anisotropy parameter $D\simeq 150$
K \cite{YO51}. This value is of importance to go further into the analysis of
the peculiar magnetic properties exhibited by this compound.

The present study also pointed to the issues that should be addressed in
priority to better describe $\chi _{\parallel }(T)$\ and $\chi _{\perp }(T)$
in Ca$_3$Co$_2$O$_6$: (i) owing to the large values of $D$, it is necessary
to develop a reliable model for $T<D$; (ii) since the intrachain coupling
seems to have an impact even at large $T$ (when considering a MF\ approach),
it would be valuable to consider a modelization accounting for the 1D\
character of this compound. These developments are presently under way.

\end{document}